\documentclass{eptcs}
\providecommand{\event}{TFPIE 2023}

\usepackage{amsmath,amssymb}
\usepackage{fontaxes}
\usepackage{graphicx}
\usepackage{stfloats}
\usepackage{multido}
\usepackage{setspace}
\usepackage{textcomp}
\usepackage{alltt}
\usepackage{textcomp}
\usepackage{pgfplots}
\pgfplotsset{compat=newest}
\usepackage{pgfplotstable}
\usepackage{cleveref}
\usepackage{doi}
\usepackage{float}
\floatstyle{ruled}
\restylefloat{figure}

\newcommand{\racket}{\texttt{Racket}}
\newcommand{\arrow}{\(\rightarrow\)}

\newcommand{\elist}{\textquotesingle{()}}
\newcommand{\fsm}{\texttt{FSM}}
\newcommand{\sig}{\texttt{\(\Sigma\)}}
\newcommand{\sigsig}{\texttt{\(\Sigma\) = \{a b\}}}
\newcommand{\lamb}{\texttt{$\lambda$}}
\newcommand{\ep}{\texttt{\(\epsilon\)}}
\newcommand{\dotss}{\(\ldots\)}

\newcommand{\quot}{\texttt{\textquotesingle{}}}
\newcommand{\dquot}{\texttt{"}}

\title{Regular Expressions in a CS Formal Languages Course}

\author{Marco T. Moraz\'an
\institute{Seton Hall University}
\email{morazanm@shu.edu}}
\def\titlerunning{Regular Expressions in a CS Formal Languages Course}
\def\authorrunning{M. T. Moraz\'an}

\begin{document}

\maketitle

\begin{abstract}
Regular expressions in an Automata Theory and Formal Languages course are mostly treated as a theoretical topic. That is, to some degree their mathematical properties and their role to describe languages is discussed. This approach fails to capture the interest of most Computer Science students. It is a missed opportunity to engage Computer Science students that are far more motivated by practical applications of theory. To this end, regular expressions may be discussed as the description of an algorithm to generate words in a language that is easily programmed. This article describes a programming-based methodology to introduce students to regular expressions in an Automata Theory and Formal Languages course. The language of instruction is \fsm{} in which there is a regular expression type. Thus, facilitating the study of regular expressions and of algorithms based on regular expressions.
\end{abstract}

\section{Introduction}

Historically, formal languages and automata theory courses are theoretical pencil-and-paper courses. Students design algorithms in theory (i.e., without implementing them) and write theorems based on the algorithms they design. If there is a bug in the algorithm then it is commonly the case (especially among students) that the bug is not discovered and there is, of course, also a bug in the proof of a theorem. This truly goes against the grain of a Computer Science education. Computer Science students are trained to design \underline{and} implement algorithms. Unit testing and runtime bugs give them immediate feedback providing the opportunity to make corrections before submitting work for grading. This is rather difficult to do if algorithms are only designed and never implemented, because few instructors, if any, have time to provide feedback on draft solutions.

Regular expressions are introduced as a finite language-representation. Such a representation is needed, because many interesting languages are not finite. That is, they contain an infinite number of words and, therefore, it is impossible to list all the words in the language. A finite representation for a language, of course, must be written using a finite number of symbols and must be different from the representation used for any other language. If \texttt{$\Sigma$} is an alphabet used to write the finite representations of languages then all possible finite language representations are in \texttt{$\Sigma^{\texttt{*}}$}. This means that the language of finite language representations is countably infinite (i.e., they can be printed in alphabetical order like the words in a complete English dictionary).

Before writing their first regular expression, an early lesson students learn  is that, \texttt{2$^{\Sigma^{\texttt{*}}}$}, $\Sigma^{\texttt{*}}$'s power set is uncountable. So, there is a countable number of finite language representations and an uncountable number of languages to represent. Therefore, a finite representation for each language does not exist. The best that we can achieve is to develop a finite representation for some interesting languages. As long as a representation is finite the majority of languages cannot be represented. Although it is important for students to understand this result, the tone set by this approach is one that is usually found too theoretical by Computer Science students in a Formal Languages and Automata Theory course. This leads to apathy towards the material. This apathy is an unfortunate side-effect because regular expressions are about programming and algorithms--a fact lost by most students studying regular expressions for the first time in a formal languages course.

This article outlines a programming-based approach to teaching students about regular expressions in their first automata theory course using \fsm{} (\emph{\texttt{F}}inite \emph{\texttt{S}}tate \emph{\texttt{M}}achines). \fsm{} is a domain-specific language in which regular expressions, state machines, and grammars are types \cite{fsm}. Students are exposed to all the theory addressed by a traditional non-programming-based automata theory course, but are engaged by programming regular expressions and by designing/implementing programs based on regular expressions. The climax of this module brings students to the realization that regular expressions are an elegant way to describe an algorithm for generating members of a language. This module is covered before finite-state automatons are discussed in a course for third- and fourth-year undergraduate students at Seton Hall University.

The article is organized as follows. \Cref{rw} reviews and contrasts approaches to teaching students about regular expressions. \Cref{regexp-fsm} briefly outlines regular expressions in \fsm. \Cref{regexp-prog} discusses classroom examples of programming with regular expressions. \Cref{wordgen} discusses how students implement a word-generating function given a regular expression. \Cref{app} discusses applications of regular expressions and outlines password generation using regular expressions. Finally, \Cref{concls} presents concluding remarks and directions for future work.

\section{Related Work}
\label{rw}

There is wide-ranging treatment of regular expressions in Formal Languages and Automata Theory textbooks. Most textbooks start with finite-state automatons and discussion leads to regular expressions (e.g., \cite{Hopcroft,Linz,Martin,Rich,Sipser}). Other textbooks start with regular expressions and then move to finite-state automatons (e.g., \cite{Lewis}). Regardless of when regular expressions are introduced, the depth of their treatment varies a great deal. Some textbooks only provide an informal definition for regular expressions, briefly discuss an application (e.g., lexical analysis), and quickly move to the equivalence between regular expressions and finite-state automatons (e.g., \cite{Martin}). Most textbooks go a little further providing a formal definition and a set of examples before moving on to the equivalence between regular expressions and finite-state automatons (e.g., \cite{Hopcroft,Linz}). The work presented in this article is delivered before finite-state automatons are presented. It provides students with a formal definition and regular expression examples. In contrast, however, the formal definition is as a type instance in a programming language and the examples are executable programs.

Sipser \cite{Sipser} and Lewis and Papadimitriou \cite{Lewis} have a more in-depth treatment of regular expressions. They motivate regular expressions as a finite representation that may be used to describe infinite languages. They provide a formal definition, examples, and discuss properties of regular expressions. Sipser briefly discusses identity properties and then lexical analysis before moving to the equivalence of regular expressions and finite-state automatons. Lewis and Papadimitriou focus a bit less on mathematical properties and informally discuss how regular expressions outline the steps to generate words in the language they describe. They reject, however, calling these steps an algorithm because of their nondeterministic nature (e.g., choosing the number of repetitions when generating a word for a Kleene star regular expression). In a similar manner, the work presented in this article presents students with a formal definition and examples. In contrast, mathematical properties of regular expressions are less emphasized and examples purposely lead to an algorithm and its implementation for generating words in a regular expression's language. The algorithm fully recognizes that randomness (i.e., nondeterminism) has its role in computation. As pointed out by Lewis and Papadimitriou, the result of generating a word is not predictable. This, however, is only part of the story. There are properties that every word generated, given a regular expression, must satisfy. Therefore, unit tests using property-based testing may be written to validate any generated word.

Rich \cite{Rich} discusses mathematical properties and several applications of regular expressions (e.g., lexical analysis, spam filtering, and password validity). Of the classical Formal Languages and Automata Theory textbooks, this is the most algorithmic. It emphasizes algorithms based on regular expressions, but only presents them as pseudocode. There is a discussion addressing the generation of words in the language of a given regular expression. For example, Rich suggests to \emph{think of any expression that is enclosed in a Kleene star as a loop that can be executed zero or more times}. As Rich, the work presented in this article focuses on algorithms. In contrast, however, the work presented in this article also focuses on implementation. A word-generating function is fully implemented based on the experience students gain from implementing regular expressions. Students walk away understanding how to design and implement a word-generating function for any given regular expression.

\section{A Brief Introduction to Regular Expressions in \fsm}
\label{regexp-fsm}

\fsm{} is a domain-specific language embedded in \racket{} \cite{Racket} and its use is specified as follows: \texttt{\#lang fsm}. It inherits from \racket{} its syntax and its rich set of primitive functions. It is a programming language designed for the Automata Theory and Formal Languages classroom. Among its types are regular expressions, finite-state machine (e.g., finite-state automata, pushdown automata, and Turing machine), and grammars (e.g., regular grammar, context-free grammar, and context-sensitive grammar). It also defines constants like \texttt{EMP} that denotes the empty word (i.e., a word of length 0). In contrast, a nonempty word is a \texttt{(listof symbol)}. Each symbol in a nonempty word is a member of an alphabet, \sig{}, that may contain symbols representing lowercase letters in the Roman alphabet, numbers, or special characters like \texttt{\&}, \texttt{!}, \texttt{\$}, and \texttt{*}.

An important feature of \fsm{} is that nondeterminism is a feature. That is, programmers are not burdened with implementing nondeterminism. Instead, nondeterminism is built into language primitives.

\subsection{Regular Expressions}

A regular expression, over an alphabet \sig, is an \fsm \ type defined as follows\footnote{The regular expression for the empty language, \texttt{(null-regexp)}, is not introduced until needed later in the course when transforming a nondeterministic finite-state automaton into a regular expression. In this manner, discussions about how operations are performed with \texttt{(null-regexp)} (e.g., \texttt{(concat A \texttt{(null-regexp)})}) are postponed until students are more familiar with regular expressions.}:
\begin{enumerate}
    \item (empty-regexp)
    \item (singleton-regexp \dquot{}a\dquot{})\textrm{, where a\(\in\)\(\Sigma\)}
    \item (union-regexp r1 r2)\textrm{,  where r1 and r2 are regular expressions}
    \item (concat-regexp r1 r2)\textrm{, where r1 and r2 are regular expressions}
    \item (kleenestar-regexp r)\textrm{, where r is a regular expression}
\end{enumerate}
Each regular expression subtype is built using a distinct constructor. The language of a regular expression, \texttt{r}, is denoted by \texttt{L(r)}. It contains all the words that can be generated with \texttt{r}. A language that is described by a regular expression is called a \emph{regular language}.

The first regular expression describes the following language:
\begin{center}
     L((empty-regexp)) = \{EMP\} = \{\ep\}
\end{center}
That is, it is a language that only contains the empty word.

The constructor \texttt{singleton-regexp} is used to build a regular expression for any element in \sig. It takes as input a string representing an element in \sig. A singleton regular expression describes the following language:
\begin{center}
     L((singleton-regexp "a")) = \{a\}
\end{center}
That is, it is the language that only contains \texttt{a}.

If \texttt{r1} and \texttt{r2} are regular expressions then a regular expression for \texttt{L(r1) $\cup$ L(r2)} is built using \texttt{union-regexp}. It describes the following language:
\begin{center}
     L((union-regexp r1 r2)) = \{w $|$ w\(\in\)L(r1) \(\vee\) w\(\in\)L(r2)\}
\end{center}
That is, it represents the language that contains all the words in \texttt{L(r1)} and all the words in \texttt{L(r2)}.

If \texttt{r1} and \texttt{r2} are regular expressions then \texttt{concat-regexp} builds a regular expression for \texttt{L(r1)L(r2)}. It describes the following language:
\begin{center}
     L((concat-regexp r1 r2)) = \{w\(\sb{1}\)w\(\sb{2}\) $|$ w\(\sb{1}\)\(\in\)L(r1) \(\wedge\) w\(\sb{2}\)\(\in\)L(r2)\}
\end{center}
That is, it is the language that contains all words constructed by concatenating a word in \texttt{L(r1)} and a word in \texttt{L(r2)}.

If \texttt{r} is a regular expression then a regular expression for zero or more concatenations of words in \texttt{L(r)} is built using \texttt{kleenestar-regexp}. It represents the following language:
\begin{center}
  L((kleenestar-regexp r)) = \{\{EMP\} \(\cup\) \{w\(\sb{1}\)w\(\sb{2}\) \dotss w\(\sb{n}\) $|$ w\(\sb{1}\),w\(\sb{2}\),\dotss,w\(\sb{n}\in\)L(r) \(\wedge\) n \(\geq\) 1\}\}
\end{center}
That is, it is the language that contains all words constructed by concatenating zero or more words in \texttt{L(r)}.

Students are made aware that \fsm{} provides informative error messages to help overcome the misuse of regular expression constructors\cite{fsmerrors}. When a constructor is misused an error is thrown. This is a sampling of \fsm \ error messages for misuse of regular expression constructors:
\begin{alltt}
     > (union-regexp 2 (singleton-regexp \textquotesingle{}w))
     \textcolor{red}{the input to the regexp #(struct:singleton-regexp w) must be a string}
     > (union-regexp (empty-regexp) 3)
     \textcolor{red}{3 must be a regexp to be a valid second input to}
     \textcolor{red}{union-regexp #(struct:empty-regexp) 3}
     > (concat-regexp 3 (empty-regexp))
     \textcolor{red}{3 must be a regexp to be a valid first input to}
     \textcolor{red}{concat-regexp 3 #(struct:empty-regexp)}
     > (kleenestar-regexp "A U B")
     \textcolor{red}{"A U B" must be a regexp to be a valid input to kleenestar-regexp}
     > (singleton-regexp 1)
     \textcolor{red}{the input to the regexp #(struct:singleton-regexp 1) must be a string}
\end{alltt}
Observe that the error messages are not prescriptive. This is important because it is impossible to discern the intention of the programmer \cite{Marceau,Mind}. Proposing a solution to the bug may lead the programmer down the wrong path to solve it in a manner consistent with her design.

\fsm{} does not burden programmers with implementing nondeterminism for regular expressions. Nondeterminism in \fsm{} regular expressions is a language feature that students do not need to know how to implement (much like they do not need to know how to implement features in their favorite programming language). To this end, the following primitives are provided:
\begin{description}
  \item[\underline{(pick-regexp r)}:] Nondeterministically returns a nested sub-\texttt{regexp} from the given \texttt{union-regexp}. This includes any nested \texttt{union-regexp}s in a chain of \texttt{union-regexp}s. For example, if the \texttt{union-regexp} is \texttt{(union-regexp r1 (union-regexp r2 (union-regexp r3 r4)))} then the selected \texttt{regexp} may be any of \texttt{r1}--\texttt{r4}.
  \item[\underline{(pick-reps n)}:] Nondeterministically generate a natural number in [0..n].
  \item[\underline{(gen-regexp-word r)}:] Nondeterministically generates a word in the language of the given \texttt{regexp}.
\end{description}
For convenience, \fsm{} also provides the following function to generate a word from a \texttt{singleton-regexp}:
\begin{description}
     \item[\underline{(convert-singleton r)}:] Converts the given \texttt{singleton-regexp} to a word of length 1.
\end{description}

Finally, \fsm{} addresses the printed representation of regular expressions. By ``printable'' we mean a string fit for humans to read. The following table outlines the printable forms of regular expressions:
\begin{center}
\begin{tabular}{|r|l|}
  \hline
  Regular Expression    & Printable Form \\ \hline
  \texttt{(empty-regexp)}        & \dquot$\epsilon$\dquot \\ \hline
  \texttt{(singleton-regexp a)}  & \dquot{}a\dquot \\ \hline
  \texttt{(union-regexp r1 r2)}  & (string-append (printable-regexp r1)\\
                                 & \ \ \ \ \ \ \ \ \ \ \ \ \ \ \ \ \ \ \ \ \ \ \ \ \dquot{}U\dquot{}\\
                                 & \ \ \ \ \ \ \ \ \ \ \ \ \ \ \ \ \ \ \ \ \ \ \ \ (printable-regexp r2))\\ \hline
  \texttt{(concat-regexp r1 r2)} & (string-append (printable-regexp r1)\\
                        & \ \ \ \ \ \ \ \ \ \ \ \ \ \ \ \ \ \ \ \ \ \ \ \ (printable-regexp r2)) \\ \hline
  \texttt{(kleenestar-regexp r)} & (string-append (printable-regexp r1) \dquot{}*\dquot{}) \\ \hline
\end{tabular}
\end{center}
The \fsm \ function \texttt{printable-regexp} returns a string representing the regular expression it is given as input. The following interactions illustrate how \texttt{printable-regexp} works:
\begin{alltt}
     > (printable-regexp (empty-regexp))
     "\(\epsilon\)"
     > (printable-regexp (singleton-regexp "z"))
     "z"
     > (printable-regexp (union-regexp
                           (singleton-regexp "z")
                           (union-regexp (singleton-regexp "1")
                                         (singleton-regexp "q"))))
     "(z U (1 U q))"
     > (printable-regexp (concat-regexp (singleton-regexp "i")
                                        (singleton-regexp "i")))
     "ii"
     > (printable-regexp (kleenestar-regexp
                           (concat-regexp (singleton-regexp "a")
                                          (singleton-regexp "b"))))

     "(ab)\(\sp{*}\)"
\end{alltt}

\subsection{Regular Expression Selectors and Predicates}
\label{ftemplate}
The \fsm \ selector functions for regular expressions are:
\begin{quote}
\begin{description}
  \item[\underline{singleton-regexp-a}:] Extracts the embedded string
  \item[\underline{kleenestar-regexp-r1}:] Extracts the embedded regular expression
  \item[\underline{union-regexp-r1}:] Extracts the first embedded regular expression
  \item[\underline{union-regexp-r2}:] Extracts the second embedded regular expression
  \item[\underline{concat-regexp-r1}:] Extracts the first embedded regular expression
  \item[\underline{concat-regexp-r2}:] Extracts the second embedded regular expression
\end{description}
\end{quote}
The following predicates are defined to distinguish among the regular expression subtypes:
\begin{alltt}
     empty-regexp?     singleton-regexp?     kleenestar-regexp?
     union-regexp?     concat-regexp?
\end{alltt}
Each consumes one value of any type and returns a Boolean. They return true if the input is a regular expression of the subtype tested. Otherwise, they return false.

Armed with the constructors, selectors, and predicates for regular expressions we can write a template for functions on a regular expression:
\begin{alltt}
     ;; regexp \dotss \arrow \dotss
     ;; Purpose: \dotss
     (define (f-on-regexp rexp \dotss)
       (cond [(empty-regexp? rexp) \dotss]
             [(singleton-regexp? rexp) \dotss(singleton-regexp-a rexp)\dotss]
             [(kleenestar-regexp? rexp)
              \dotss(f-on-regexp (kleenestar-regexp-r1 rexp))\dotss]
             [(union-regexp? rexp)
              \dotss(f-on-regexp (union-regexp-r1 rexp))\dotss
              \dotss(f-on-regexp (union-regexp-r2 rexp))\dotss]
             [else \dotss(f-on-regexp (concat-regexp-r1 rexp))\dotss
                   \dotss(f-on-regexp (concat-regexp-r2 rexp))\dotss]))
\end{alltt}
The function template reflects the structure of regular expressions and suggests using structural recursion to process a regular expression.

\section{Programming with Regular Expressions}
\label{regexp-prog}

Given that \fsm{} has a regular expression type, their design may follow a top-down or a bottom-up divide-and-conquer approach. The idea is to define a regular expression by parts. Just like a program is composed of one or more functions, a regular expression is composed of one or more regular expressions. In this section we explore how to design and implement regular expressions.

\subsection{All Words Ending with an \texttt{a}}
\label{ends-a}

The first (gentle) exercise building a finite representation for an infinite language starts with the following language over \sigsig:
\begin{center}
     L = \{w $|$ w ends with an a\}
\end{center}
Students are asked if a regular expression for \texttt{L} can be programmed. To start, a top-down design is followed. Every word in \texttt{L} must have at least one \texttt{a} at the end. Before the last \texttt{a} there can be an arbitrary number of \texttt{a}s and \texttt{b}s. Assuming a regular expression can be developed for both parts, the regular expression for \texttt{L} that concatenates the languages for each part may be written as follows:
\begin{alltt}
     (define ENDS-WITH-A  (concat-regexp AUB* A))
\end{alltt}
The regular expression \texttt{A} must represent \texttt{a}. This is defined as follows:
\begin{alltt}
     (define A (singleton-regexp "a"))
\end{alltt}
\texttt{AUB*} must represent an arbitrary number of elements. This suggests defining a \texttt{kleenestar-regexp}. This may be done as follows:
\begin{alltt}
     (define AUB* (kleenestar-regexp AUB))
\end{alltt}
\texttt{AUB} represents a choice between a word in \texttt{L(A)} and a word in \texttt{L(B)}. Having a choice suggests defining a \texttt{union-regexp}:
\begin{alltt}
     (define AUB (union-regexp A B))
\end{alltt}
Finally, \texttt{B} represents the singleton regular expression for \texttt{"b"}. This is done as follows:
\begin{alltt}
     (define B (singleton-regexp "b"))
\end{alltt}
The complete program for a regular expression for \texttt{L} is displayed in \Cref{endsa}. The unit test is written using \texttt{check-equal?} provided by, \texttt{rackunit}, \racket's unit-testing framework.

\begin{figure}[t!]
\begin{alltt}
     (define A (singleton-regexp "a"))

     (define B (singleton-regexp "b"))

     (define AUB (union-regexp A B))

     (define AUB* (kleenestar-regexp AUB))

     (define ENDS-WITH-A  (concat-regexp AUB* A))

     (check-equal? (printable-regexp ENDS-WITH-A) "(a U b)*a")
\end{alltt}
\caption{The \fsm \ program for L = \{w $|$ w ends with an \texttt{a}\}.} \label{endsa}
\end{figure}

\subsection{Binary Numbers}

Next students are asked to consider a slightly more complex language that ought to be, at least informally, familiar to them. They consider the following language:
\begin{center}
     BIN-NUMS = \{w $|$ w is a binary number without leading zeroes\}
\end{center}
Although the above definition may sound clear to students, it is lacking. It does not provide any details about the structure of the binary numbers in the language nor any indication on how to build such numbers. Class turns to formally define BIN-NUMS using a regular expressions. To provide a different development perspective, a bottom-up divide-and-conquer approach is used.

\subsubsection{Implementing a Regular Expression}

Based on the problem statement the following observations are made:
\begin{center}
\begin{minipage}[c]{0.9\linewidth}
\begin{enumerate}
  \item \texttt{$\Sigma$ = \{0 1\}}
  \item The minimum length of a binary number is 1
  \item A binary number with a length greater than 1 cannot start with 0
\end{enumerate}
\end{minipage}
\end{center}
The second observation informs us that the empty regular expression is not part of the language. The third observation informs us that there are two subtypes of binary numbers in the set: \texttt{0} and those starting with 1 followed by an arbitrary number of \texttt{0}s and \texttt{1}s.

\begin{figure}[t!]
\begin{alltt}
     (define ZERO (singleton-regexp "0"))

     (define ONE  (singleton-regexp "1"))

     (define 0U1* (kleenestar-regexp (union-regexp ZERO ONE)))

     (define STARTS1 (concat-regexp ONE 0U1*))

     (define BIN-NUMS (union-regexp ZERO STARTS1))

     (check-equal? (printable-regexp BIN-NUMS) "(0 U 1(0 U 1)*)")
\end{alltt}
\caption{A program defining binary numbers without leading zeroes} \label{bin-nums}
\end{figure}

The simplest regular expressions needed are for the elements of \texttt{$\Sigma$}. These are all singleton regular expressions:
\begin{alltt}
  (define ZERO (singleton-regexp "0"))   (define ONE  (singleton-regexp "1"))
\end{alltt}
An arbitrary number of \texttt{0}s and \texttt{1}s may be represented using a union and a Kleene star regular expression:
\begin{alltt}
     (define 0U1* (kleenestar-regexp (union-regexp ZERO ONE)))
\end{alltt}
With the above definition, a regular expression for 1 followed by an arbitrary number of \texttt{0}s and \texttt{1}s is:
\begin{alltt}
     (define STARTS1 (concat-regexp ONE 0U1*))
\end{alltt}
Finally, \texttt{BIN-NUMS} may be implemented by a union regular expression to provide a choice among the subtypes. \Cref{bin-nums} displays the complete program to define \texttt{BIN-NUMS}.

\subsubsection{Generating \texttt{BIN-NUMS} Words}
\label{binary-nums}

Students are asked to compare the two formulations for \texttt{BIN-NUMS} and determine which is more useful:
\begin{itemize}
     \item BIN-NUMS = \{w $|$ w is a binary number without leading zeroes\}

     \item BIN-NUMS = (0 \(\cup\) 1(0 \(\cup\) 1)\(\sp{*}\))
\end{itemize}
The truth is that both are useful. The first provides a quick intuitive understanding of what the language represents. It lacks, however, any description for constructing words. In this regard, the second formulation is more useful. It describes an algorithm for constructing binary numbers without leading zeroes. Either generate 0 or generate 1 followed by an arbitrary number of \texttt{0}s or \texttt{1}s.

This means that a function to generate a random word in \texttt{BIN-NUMS} can and ought to be implemented. We shall follow the steps of the design recipe for a function to write this program \cite{HtDP2,APS}. To simplify the discussion the default number of Kleene star repetitions shall be 10. Based on this design idea, the next steps of the design recipe are outlined as follows:
\begin{alltt}
     ;; [natnum>0] \arrow BIN-NUMS
     ;; Purpose: Generate a binary number without leading zeroes
     (define (generate-bn . n)

       (define MAX-KS-REPS (if (null? n) 10 (first n)))

       ;; regexp \arrow word
       ;; Purpose: Generate a word representing a valid binary number,
       ;;          such that the number of Kleene star repetitions is
       ;;          in [0..MAX-KS-REPS].
       (define (gen-word r) \dotss)

      (gen-word BIN-NUMS))
\end{alltt}
The signature and function header define, respectively, a single optional argument for the maximum number of repetitions for a Kleene star regular expression. The purpose statement clearly and briefly describes the problem solved. A local constant is defined for the Kleene star repetitions with 10 as the default value. Finally, the local function \texttt{gen-word} generates a binary number and is designed by specializing the template for functions on a regular expression.

The next step of the design recipe is to write unit tests. We are unable, however, to write tests using \texttt{check-equals?} because the program randomly generates elements in \texttt{BIN-NUMS}. To test such a function we use property-based testing. That is, we shall test that the generated words have the expected properties. To write tests we use \texttt{rackunit}'s \texttt{check-pred}. \texttt{check-pred} requires a predicate to test and input for the predicate. If the predicate holds the test passes. If the predicate does not hold the test fails and a failed test report is generated. Any word, \texttt{w}, generated by \texttt{generate-bn} must have the following properties:
\begin{center}
\begin{minipage}[c]{0.9\linewidth}
\begin{enumerate}
  \item \texttt{w} is a list (i.e., it cannot be \texttt{EMP})
  \item 1 \texttt{$\leq$} \texttt{(length w)}
  \item \texttt{w} is \texttt{\quot(0)} or \texttt{(first w)} is 1
  \item \texttt{w} only contains \texttt{0}s and \texttt{1}s
\end{enumerate}
\end{minipage}
\end{center}
To test that a generated word represents a binary number we ignore the length limit. Following the steps of the design recipe yields the following predicate:
\begin{alltt}
   ;; word \arrow Boolean
   ;; Purpose: Test if the given word is in L(BIN-NUMS)
   (define (is-bin-nums? w)
     (and (list? w)
          (<= 1 (length w))
          (or (equal? w \quot(0)) (= (first w) 1))
          (andmap (\lamb (bit) (or (= bit 0) (= bit 1))) w)))

   (check-equal? (is-bin-nums? \quot()) #f)
   (check-equal? (is-bin-nums? \quot(0 0 0 1 1 0 1 0)) #f)
   (check-equal? (is-bin-nums? \quot(0)) #t)
   (check-equal? (is-bin-nums? \quot(1 0 0 1 0 1 1)) #t)
   (check-equal? (is-bin-nums? \quot(1 1 1 0 1 0 0 0 1 1 0 1)) #t)

\end{alltt}
This predicate is used to write the tests for \texttt{generate-bn}. Anything returned by \texttt{generate-bn} must satisfy \texttt{is-bin-nums?}. The tests are:
\begin{alltt}
   (check-pred is-bin-nums? (generate-bn))
   (check-pred is-bin-nums? (generate-bn))
   (check-pred is-bin-nums? (generate-bn))
   (check-pred is-bin-nums? (generate-bn))
   (check-pred is-bin-nums? (generate-bn))
\end{alltt}
Although the tests all look the same they are not the same test. Recall that \texttt{(generate-bn)} is nondeterministic (i.e., the output value cannot be predicted). Therefore, each test above is for a different word (not necessarily distinct) returned by \texttt{(generate-bn)}.

\begin{figure}[t!]
\begin{alltt}
     #lang fsm
     ;; [natnum>0] \arrow word
     ;; Purpose: Generate a binary number without leading zeroes
     ;;          unless its 0
     (define (generate-bn . n)
       (define MAX-KS-REPS (if (null? n) 10 (first n)))

       ;; regexp \arrow{} word
       ;; Purpose: Generate a word representing a valid binary number,
       ;;          such that the number of Kleene star repetitions is
       ;;          in [0..MAX-KS-REPS]
       (define (gen-word r)
         (cond [(singleton-regexp? r) (convert-singleton r)]
               [(concat-regexp? r)
                (let [(w1 (gen-word (concat-regexp-r1 r)))
                      (w2 (gen-word (concat-regexp-r2 r)))]
                  (append w1 w2))]
               [(union-regexp? r) (gen-word (pick-regexp r))]
               [(kleenestar-regexp? r)
                (flatten (build-list
                           (pick-reps MAX-KS-REPS)
                           (\lamb{} (i) (gen-word (kleenestar-regexp-r1 r)))))])
        (gen-word BIN-NUMS))
\end{alltt}
\caption{The function to generate words in \texttt{BIN-NUMS}.} \label{gen-binnums}
\end{figure}

The function \texttt{gen-word} must be able to process any \texttt{regexp} subtype that is part of the \texttt{BIN-NUMS}: \texttt{singleton-regexp}, \texttt{concat-regexp}, \texttt{union-regexp}, and \texttt{kleenestar-regexp}. The function template to process a \texttt{regexp} is specialized. To process a \texttt{singleton-regexp}, \texttt{convert-singleton} is used. To process a \texttt{concat-regexp}, two words are generated, using each embedded \texttt{regexp}, and appended. To process a \texttt{union-regexp}, a recursive call is made with one of the sub-\texttt{regexps} nondeterministically chosen by \texttt{pick-regexp}.  To process a \texttt{kleenestar-regexp}, the number of binary numbers to generate is nondeterministically chosen using \texttt{pick-reps}. A list containing that number of binary numbers is generated and flattened. The result of this design is displayed in \Cref{gen-binnums}.

\section{Generating Words in the Language Defined by a Regular Expression}
\label{wordgen}

The development of \texttt{generate-bn} confirms that a regular expression, \texttt{r}, describes a construction algorithm for words in \texttt{L(r)}. This suggests that the algorithm can be generalized to generate an arbitrary word in the language of an arbitrary regular expression.

\subsection{Design Idea}

The function takes as input a regular expression and an optional natural number for the maximum number of Kleene star repetitions. It returns a word. A constant, \texttt{MAX-KLEENESTAR-REPS}, is locally defined for the maximum number of Kleene star repetitions. If the optional natural number is not provided the default value of the constant is arbitrarily defined to be 20. Building on the experience with generating binary numbers, there is a local function, \texttt{generate}, to generate a word.

As suggested by the function template to process a \texttt{regexp}, \texttt{generate} must distinguish among the regular expression subtypes to generate the word. If the input is the empty regular expression then the only word that may be generated is, \texttt{EMP}, the empty word. If given a singleton regular expression then \texttt{convert-singleton} is used to generate a word of length 1 from the embedded string.

To process a Kleene star regular expression, a list of words is generated using the embedded \texttt{regexp}. The length of the list is nondeterministically chosen using \texttt{pick-reps}. Once generated, the list of words is filtered for empty words and flattened. If the resulting list is empty then \texttt{EMP} is returned. Otherwise, the resulting list is returned.

To process a union regular expression, \texttt{pick-regexp} is used to nondeterministically select one of the expressions in the union. A recursive call is made with the selected regular expression to generate the word.

To process a \texttt{concat-regexp}, a word is generated using each of the embedded regular expressions. If both generated words are \texttt{EMP} then \texttt{EMP} is returned. If either word is \texttt{EMP} then the other word is returned. Otherwise, the two generated words are appended and returned.

\subsection{Signature, Purpose, and Function Header}

Students are reminded that the signature, purpose statement, and function header collectively provide documentation that explains to any reader of the code what the function is expected to do. The next steps of the design recipe are satisfied as follows:
\begin{alltt}
     ;; regexp [natnum] \arrow word
     ;; Purpose: Generate a random word in the language
     ;;          of the given regexp such that the number
     ;;          of repetitions generated from a Kleene
     ;;          star regular expression does not exceed
     ;;          the given optional number or, otherwise,
     ;;          20.
     (define (gen-word rexp . reps)

\end{alltt}

\subsection{Tests}

To simplify the development of tests we use \texttt{ENDS-WITH-A} from \Cref{endsa} and \texttt{BIN-NUMS} from \Cref{bin-nums}. Given that we are designing a nondeterministic function, property-based testing is employed. This means we must design and implement a predicate for words ending with an \texttt{a} just like \texttt{is-bin-nums?} is designed for \texttt{BIN-NUMS}. If the given word is a list, has a length greater than or equal to 1, and its last element is an \texttt{a} then it is a word in \texttt{L(ENDS-WITH-A)}. Following the steps of the design recipe yields this predicate:
\begin{alltt}
     ;; word \arrow Boolean
     ;; Purpose: Test if the given word is in ENDS-WITH-A
     (define (is-ends-with-a? w)
       (and (list? w) (>= (length w) 1) (eq? (last w) \quot{}a)))

     (check-equal? (is-ends-with-a? \quot(a)) #t)
     (check-equal? (is-ends-with-a? \quot(b b a)) #t)
     (check-equal? (is-ends-with-a? \quot(a b b a b a)) #t)
     (check-equal? (is-ends-with-a? \quot()) #f)
     (check-equal? (is-ends-with-a? \quot(b b b)) #f)
     (check-equal? (is-ends-with-a? \quot(a a a a b)) #f)
\end{alltt}

The tests for \texttt{gen-word} are:
\begin{alltt}
     (check-pred is-bin-nums? (gen-word BIN-NUMS))
     (check-pred is-bin-nums? (gen-word BIN-NUMS))
     (check-pred is-bin-nums? (gen-word BIN-NUMS))
     (check-pred is-bin-nums? (gen-word BIN-NUMS 30))
     (check-pred is-bin-nums? (gen-word BIN-NUMS 50))

     (check-pred is-ends-with-a? (gen-word ENDS-WITH-A))
     (check-pred is-ends-with-a? (gen-word ENDS-WITH-A))
     (check-pred is-ends-with-a? (gen-word ENDS-WITH-A 18))
     (check-pred is-ends-with-a? (gen-word ENDS-WITH-A 7))
\end{alltt}

\subsection{Function Body}

The next step of the design requires writing the function's body. This is done by specializing the \texttt{cond}-expression in the template for functions on a regular expression. We independently present the design of each stanza.

For the empty regular expression the only word that can be generated is \texttt{EMP}. The corresponding stanza is:
\begin{alltt}
     [(empty-regexp? rexp) EMP]
\end{alltt}

For a \texttt{singleton-regexp}, a word is generated using \texttt{convert-singleton}:
\begin{alltt}
     [(singleton-regexp? rexp) (convert-singleton rexp)]
\end{alltt}

For a \texttt{kleenestar-regexp}, the length of the word is nondeterministically chosen using \texttt{pick-reps}. A list of words of the chosen length, generated using the embedded regular expression, is filtered to remove all \texttt{EMP}s and flattened. If the flattened list is empty then \texttt{EMP} is returned. Otherwise, the flattened list is returned. The required code is:
\begin{alltt}
     [(kleenestar-regexp? rexp)
      (let*
       [(reps (pick-reps MAX-KLEENESTAR-REPS))
        (low (flatten
              (filter
               (\lamb{} (w) (not (eq? w EMP)))
                (build-list
                 reps
                 (lambda (i)
                  (gen-word (kleenestar-regexp-r1 rexp)))))))]
       (if (empty? low) EMP low))]
\end{alltt}

For a union regular expression, a word is generated by nondeterministically picking a regular expression from the options in the union and making a recursive call with the maximum number of repetitions:
\begin{alltt}
     [(union-regexp? rexp) (gen-word (pick-regexp rexp))]
\end{alltt}

For a concatenation regular expression, two words are generated using each embedded regular expression. The words are examined as described in the design idea to return the generated word. The default stanza of the conditional is:
\begin{alltt}
   [else
    (let [(w1 (gen-word (concat-regexp-r1 rexp)))
          (w2 (gen-word (concat-regexp-r2 rexp)))]
      (cond [(and (eq? w1 EMP) (eq? w2 EMP)) EMP]
            [(eq? w1 EMP) w2]
            [(eq? w2 EMP) w1]
            [else (append w1 w2)]))]
\end{alltt}

It is suggested to students to take time to appreciate what has been achieved. A regular expression, simultaneously, is a description of, \texttt{L}, a regular language and a description of a construction algorithm for the words in \texttt{L}. It is suggested to students that a construction algorithm is an elegant way of specifying a language.

\section{Regular Expression Applications}
\label{app}
Before ending the module on regular expressions, students are introduced to one or more applications. This is something that always helps students appreciate the material they are learning. At this point, students realize that regular expressions capture a pattern for the construction of languages. As such, they are told, that regular expressions are easily found in many areas of Computer Science and, indeed, in life. It is important to note that the term regular expression is used differently in different domains. That is, a regular expression is not always defined as defined in an automata theory course. Generally, all definitions have union, concatenation, and Kleene star operations. They, however, also include other operations. These other operations may provide the ability to describe languages that are not regular. The syntax, of course, also varies. Consider, for example, the following \texttt{Perl} code snippet:
\begin{alltt}
     \$foo =\(\backsim\) m/fsm/
\end{alltt}
This expression evaluates to true if \texttt{\$foo} contains \texttt{fsm}. Put differently, it evaluates to true if \texttt{\$foo}'s value is a word in the following language:
\begin{alltt}
     L = \{w | \textit{x\textquotesingle{}},\textit{y\textquotesingle{}}\(\in\)\(\Sigma\sp{*} \wedge\) w = \textit{x\textquotesingle{}}fsm\textit{y\textquotesingle{}}\}
\end{alltt}
Clearly, \texttt{L} is a regular language. Regular expressions in \texttt{Perl}, however, are strong enough to match languages that are not regular. Therefore, students are advised that when speaking about regular expressions it is important to be precise.

Regular expressions may be used to describe, for example, internet addresses, proteins, decimal numbers, and patterns to search for in text among others. To illustrate the use of regular expressions we explore the problem of generating passwords. As always, we follow the steps of the design recipe to write a password generating function.

\subsection{Data Definitions}

A password is a string that:
\begin{itemize}
  \item  Has length $\geq$ 10
  \item  Includes at least one of each: lowercase letter, uppercase letter, and special character (i.e., \$, \&, !, or *)
\end{itemize}
Based on this definition, the sets for lowercase letters, uppercase letters, and special characters are defined as follows:
\begin{alltt}
     (define lowers \quot(a b c d e f g h i j k l m n o p q r s t u v w x y z))
     (define uppers \quot(A B C D E F G H I J K L M N O P Q R S T U V W X Y Z))
     (define spcls  \quot(\$ \& ! *))
\end{alltt}
The corresponding sets of regular expressions are defined as:
\begin{alltt}
  (define lc  (map (\lamb (lcl) (singleton-regexp (symbol->string lcl))) lowers))
  (define uc  (map (\lamb (ucl) (singleton-regexp (symbol->string ucl))) uppers))
  (define spc (map (\lamb (sc) (singleton-regexp (symbol->string sc))) spcls))
\end{alltt}

How is a password defined? To create passwords we need a regular expression. Once a password is generated it can be transformed into a string. The order in which lowercase letters, uppercase letters, and special characters appear is arbitrary. There must be, however, an, \texttt{L}, lowercase letter, a, \texttt{U}, uppercase letter, and a, \texttt{S}, special character. There are six different orderings these required elements may appear in:
\begin{alltt}
     L U S     U L S     S U L     L S U     U S L     S L U
\end{alltt}
Before and after each required element there may be an arbitrary number lowercase letters, uppercase letters, and special characters. A union regular expression is needed for the lowercase letters, for the uppercase letters, for the special characters, and for the arbitrary characters that may appear between required characters. It is emphasized to students that a union regular expression is used because it provides the ability to choose any element. These may be defined as follow:
\begin{alltt}
   (define LOWER (create-union-regexp lc))
   (define UPPER (create-union-regexp uc))
   (define SPCHS (create-union-regexp spc))
   (define ARBTRY (kleenestar-regexp
                   (union-regexp LOWER (union-regexp UPPER SPCHS))))
\end{alltt}
The creation of a chain of union regular expressions is delegated to, \texttt{create-union-regexp}, an auxiliary function (to be designed and implemented). It is now possible to define a regular expression for each of the six orderings of required elements. For instance, the regular expression for words that have the required lowercase letter first, the uppercase letter second, and the special character third is defined as follows:
\begin{alltt}
 (define LUS (concat-regexp
               ARBTRY
               (concat-regexp
                 LOWER
                 (concat-regexp
                   ARBTRY
                   (concat-regexp
                     UPPER
                     (concat-regexp ARBTRY (concat-regexp SPCHS ARBTRY)))))))
\end{alltt}
The regular expressions for the remaining 5 orderings are similarly defined.

The language of passwords is a word in any of the languages for the different orderings of required elements. It is defined using a union regular expression:
\begin{alltt}
     (define PASSWD (union-regexp
                      LUS
                      (union-regexp
                        LSU
                        (union-regexp
                          SLU (union-regexp SUL (union-regexp USL ULS))))))
\end{alltt}
Finally, in order to prevent generated passwords from getting unwieldy long \texttt{MAX-KLEENESTAR-REPS} is redefined as follows:
\begin{alltt}
     (define MAX-KLEENESTAR-REPS 5)
\end{alltt}

\subsection{Design Idea}

The constructor for a password takes no input and returns a string. A potential new password is locally defined. A word is generated by applying \texttt{gen-regexp-word} (developed in \Cref{wordgen}) to \texttt{PASSWD} and then converting the result to a string. If the length of the string is greater than or equal to 10 then it is returned as the generated password. Otherwise, a new password is generated.

\subsection{Function Definition}

Following the steps of the design recipe yields the following function definition:
\begin{alltt}
     ;;  \arrow string
     ;; Purpose: Generate a valid password
     (define (generate-password)
       (let [(new-passwd (passwd->string (gen-regexp-word PASSWD)))]
         (if (>= (string-length new-passwd) 10)
             new-passwd
             (generate-password))))
\end{alltt}

\subsection{Tests}

Given that the function is nondeterministic, property-based testing is used. For this, a predicate that takes as input a string representing a password is needed. The given password is converted into a list of symbols representing the characters in the string. This list must have a length of at least 10 and contain one element in each of the following: \texttt{lowers}, \texttt{uppers}, and \texttt{spcls}. Following the steps of the design recipe yields:
\begin{alltt}
     ;; string \arrow Boolean
     ;; Purpose: Test if the given string is a valid password
     (define (is-passwd? p)
       (let [(los (str->los p))]
         (and (>= (length los) 10)
              (ormap (\lamb (c) (member c los)) lowers)
              (ormap (\lamb (c) (member c los)) uppers)
              (ormap (\lamb (c) (member c los)) spcls))))
\end{alltt}
Converting the password to a list is delegated to, \texttt{str->los}, an auxiliary function.

Sample tests using \texttt{check-pred} are:
\begin{alltt}
     (check-pred is-passwd? (generate-password))
     (check-pred is-passwd? (generate-password))
     (check-pred is-passwd? (generate-password))
     (check-pred is-passwd? (generate-password))
     (check-pred is-passwd? (generate-password))
\end{alltt}

\subsection{Auxiliary Functions}

Three auxiliary functions are needed: \texttt{create-union-regexp}, \texttt{str->los}, and \texttt{passwd->string}. The function \texttt{create-union-regexp} takes as input a list of regular expressions and returns a union regular expression. If the length of the given list is less than 2 an error is thrown because at least two regular expressions are needed for the union of regular expressions. If the given list only has two elements then a union regular expression  is constructed with the two regular expressions in the list. If the given list has a length greater than to 2 then a union regular expression is constructed with the first regular expression in the list and the union regular expression obtained from recursively processing the rest of the list. Following the steps of the design recipe yields:
\begin{alltt}
 ;; (listof regexp) \arrow union-regexp throws error
 ;; Purpose: Create union-regexp using given list of regular expressions
 (define (create-union-regexp L)
   (cond [(< (length L) 2)
          (error "create-union-regexp: list too short")]
         [(empty? (rest (rest L))) (union-regexp (first L) (second L))]
         [else (union-regexp (first L) (create-union-regexp (rest L)))]))

 ;; Tests
 (check-equal? (create-union-regexp (list (first lc) (first uc)))
               (union-regexp (singleton-regexp "a") (singleton-regexp "A")))
 (check-equal? (create-union-regexp (list (first lc) (fourth uc) (third spc)))
               (union-regexp
                (singleton-regexp "a")
                (union-regexp (singleton-regexp "D") (singleton-regexp "!"))))
\end{alltt}

The function \texttt{str->los} takes as input a string and returns a list of symbols. The given string is converted to a list (of characters). Each character in the resulting list is converted to a symbol using \texttt{map}. The function given to \texttt{map} consumes a character and first converts the character into a string and then converts the string into a symbol. Following the steps of the design recipe yields:
\begin{alltt}
     ;; string \arrow (listof symbol)
     ;; Purpose: Convert the given string to a list of symbols
     (define (str->los str)
       (map (\lamb (c) (string->symbol (string c))) (string->list str)))

     ;; Tests
     (check-equal? (str->los "") \elist)
     (check-equal? (str->los "a!Cop") \quot(a ! C o p))
\end{alltt}

Finally, the function \texttt{passwd->string} converts a given word representing a password into a string. First, the given word is converted into a list of characters using \texttt{map}. The function given to \texttt{map} converts a symbol into a character by transforming the symbol into a string, transforming the string into a list of characters, and finally taking the first (and only) element in the list of characters. Second, the list of characters produced by \texttt{map} is converted into a string. The steps of the design recipe produce:
\begin{alltt}
     ;; word \arrow string
     ;; Purpose: Convert the given password to a string
     (define (passwd->string passwd)
       (list->string
        (map (\lamb (s) (first (string->list (symbol->string s)))) passwd)))

     ;;Tests
     (check-equal? (passwd->string \quot(a j h B ! ! y y t c)) "ajhB!!yytc")
     (check-equal? (passwd->string \quot(\$ u t q x ! J i n * K C)) "$utqx!Jin*KC")
\end{alltt}

\subsection{Running the Tests}

The students run the program and confirm that all the tests pass. In addition, students are encouraged to generate a few passwords. These are sample passwords generated:
\begin{alltt}
     > (generate-password)
     "\&\&!\$m*F!\&\$*"
     > (generate-password)
     "!e*e!*oS!lq\$"
     > (generate-password)
     "!y*\$r!C\&*d\$"
     > (generate-password)
     "\&\&!p\$rUA\$*"
     > (generate-password)
     "W\&*!eKY**D"
     > (generate-password)
     "vxY*We!Wx*\&\&u"
\end{alltt}
Students feel a sense of accomplishment seeing the results. The feeling is that passwords generated are robust and it is unlikely that anyone would be able to guess any of them.

\section{Concluding Remarks}
\label{concls}

The work presented outlines a didactic approach for introducing students to regular expressions. Unlike most Formal Languages and Automata Theory textbooks that emphasize mathematical properties to simplify regular expressions, the work presented emphasizes algorithm design and implementation. This helps keep Computer Science students motivated and engaged. Students can program regular expressions and algorithms based on regular expressions in \fsm. The importance of presenting practical and relevant applications is not ignored. Nearly all students appreciate, for example, the development of the program to generate passwords. Most students comment that it is an approach they had never had thought about before and are eager for other examples. To address this, homework exercises involving token generation or DNA pattern generation are effective.

Future work will address creating a database of examples instructors and students may draw upon for practice or presentation. The goal is to have a diverse set of examples--after all, for instance, not every student may find password generation exciting. In addition, extensions to \fsm{} are being considered. For example, it may be useful for some students to endow \fsm{} with a primitive to generate words in the language of a given regular expression.

\bibliographystyle{eptcs}
\bibliography{regexp-bib}
\end{document}